\documentclass[12pt,preprint]{aastex}
\usepackage{xspace}
\usepackage{graphicx}
\usepackage{natbib}
\newcommand{\expo}[1]{\ensuremath{10^{#1}}\xspace}

\newcommand{\gsim}{\ensuremath{\,\gtrsim\,}\xspace}
\newcommand{\lsim}{\ensuremath{\,\lesssim\,}\xspace}
\newcommand{\gl}{\ensuremath{\ell}\xspace}
\newcommand{\gb}{\ensuremath{{\it b}}\xspace}
\newcommand{\vlsr}{\ensuremath{V_{\rm LSR}}\xspace}

\newcommand{\lv}{\ensuremath{(\gl,v)}\xspace}

\newcommand{\kms}{\ensuremath{\,{\rm km\,s^{-1}}}\xspace}

\newcommand{\kmsch}{\ensuremath{\,{\rm km\,s^{-1}\,chan^{-1}}}\xspace}
\newcommand{\microns}{\ensuremath{\,\mu{\rm m}}\xspace}

\newcommand{\cm}{\ensuremath{\,{\rm cm}}\xspace}
\newcommand{\pc}{\ensuremath{\,{\rm pc}}\xspace}
\newcommand{\kpc}{\ensuremath{\,{\rm kpc}}\xspace}

\newcommand{\myr}{\ensuremath{\,{\rm Myr}}\xspace}

\newcommand{\ghz}{\ensuremath{\,{\rm GHz}}\xspace}

\newcommand{\mjy}{\ensuremath{\,{\rm mJy}}\xspace}

\newcommand{\te}{\ensuremath{{T_{e}}}\xspace}

\newcommand{\ro}  {\ensuremath{\,R_0}\xspace}         
 
\newcommand{\rgal}{\ensuremath{\,R_G}\xspace}   
\newcommand{\dsun}{\ensuremath{\,d_\sun}\xspace}          

\newcommand{\hi}{{\rm H\,{\small I}}\xspace}
\newcommand{\hii}{{\rm H\,{\small II}}\xspace}

\newcommand{\he}[1]{\ensuremath{^{#1}{\rm He}}\xspace}
\newcommand{\hal}[1]{\ensuremath{{\rm H}\,{#1}\,\alpha\xspace}}
\newcommand{\co} {\ensuremath{^{\rm 12}{\rm CO}}\xspace}

\shorttitle{H~{\small \bf II} Region Discovery Survey}
\shortauthors{Bania et al.}

\begin{document}

\title{The Green Bank Telescope Galactic H\,{\small \bf II} Region Discovery Survey}

\author{T. M. Bania\altaffilmark{1}, L. D. Anderson\altaffilmark{1,2},
Dana S. Balser\altaffilmark{3}, \& R. T. Rood\altaffilmark{4}}

\altaffiltext{1}{Astronomy Department,
725 Commonwealth Ave., Boston University, Boston MA 02215, USA.}
\altaffiltext{2}{Current Address: Laboratoire d'Astrophysique de 
Marseille (UMR 6110 CNRS \& Universit\'e de Provence), 38 rue F. 
Joliot-Curie, 13388 Marseille Cedex 13, France.}
\altaffiltext{3}{National Radio Astronomy Observatory, 520 Edgemont Road, 
Charlottesville VA, 22903-2475, USA.}
\altaffiltext{4}{Astronomy Department, University of Virginia, P.O. Box 3818, 
Charlottesville VA 22903-0818, USA.}

\begin{abstract}

We discovered a large population of previously unknown Galactic \hii
regions by using the Green Bank Telescope to detect their hydrogen
radio recombination line emission. Since recombination lines are
optically thin at 3\cm wavelength, we can detect \hii regions across
the entire Galactic disk.  Our targets were selected based on
spatially coincident 24\microns and 21\cm continuum emission.  For the
Galactic zone $-16\,\degr \leq \gl \leq 67\,\degr$ and $\vert \gb
\vert \leq 1\,\degr$ we detected 602 discrete recombination line
components from 448 lines of sight, 95\% of the sample targets, which
more than doubles the number of known \hii regions in this part of the
Milky Way.  We found 25 new first quadrant nebulae with negative LSR
velocities, placing them beyond the Solar orbit.  Because we can
detect all nebulae inside the Solar orbit that are ionized by O-stars,
the Discovery Survey targets, when combined with existing \hii region
catalogs, give a more accurate census of Galactic \hii regions and
their properties.  The distribution of \hii regions across the
Galactic disk shows strong, narrow ($\sim$\,1 kpc wide) peaks at
Galactic radii of 4.3 and 6.0 kpc.  The longitude-velocity
distribution of \hii regions now gives unambiguous evidence for
Galactic structure, including the kinematic signatures of the radial
peaks in the spatial distribution, a concentration of nebulae at the
end of the Galactic Bar, and nebulae located on the kinematic locus of
the 3\kpc Arm.

\end{abstract}


\newpage

\section{INTRODUCTION\label{sec:intro}}

\hii\ regions are the formation sites of massive OB stars.
Because the main sequence lifetimes of OB stars are $\lsim10\myr$,
\hii\ regions are zero age objects compared to
the age of the Milky Way.  They thus reveal the locations of current
Galactic star formation.  They are the archetypical tracers of Galactic
spiral structure. Their chemical abundances, moreover, indicate the
present state of the interstellar medium (ISM) and reveal the
elemental enrichment caused by the nuclear processing of many stellar
generations.  They provide unique and important probes of billions of
years of Galactic chemical evolution (GCE).
Individual Galactic \hii\ regions are astrophysically important
objects that reveal details of the impact of the star formation
process on the ISM.  Knowing the physical properties of
\hii\ region/photo-dissociation region/molecular cloud complexes
provides important constraints on the physics of star formation and
the evolution of the ISM.

Modern Galactic \hii\ region surveys began with studies of the Palomar
optical survey plates \citep{sharpless53, sharpless59}.  Radio
recombination lines (RRLs) from Galactic \hii regions were discovered
in 1965 during the commissioning of the NRAO 140 Foot telescope;
\citet{hoglund65} detected \hal{109} emission at 6\cm wavelength from
M\,17 and Ori\,A.  Because the Galactic ISM is optically thin at
centimeter wavelengths, RRL surveys were able to discover large
numbers of \hii regions distributed throughout the entire Galactic
disk.  Pioneering RRL surveys were done by, e.g., Reifenstein et
al. 1970, Wilson et al. 1970, Downes et al. 1980, Caswell \& Haynes
1987, and Lockman 1989.  These surveys gave important insights into
Galactic structure and the spatial pattern of massive star formation.
Particularly noteworthy was the discovery of a metallicity gradient
across the Galactic disk, made apparent by RRL measurements of \hii
region electron temperatures \citep{wink83, shaver83, quireza06}.
This Galactic-scale metallicity gradient placed strong constraints on
GCE.
By the time of the \citet{lockman89} survey, however, almost all of
the reasonably strong radio continuum sources, as revealed by contour
maps often drawn by hand, had been observed; large angular scale
surveys for discrete \hii regions using \cm-wave RRLs as tracers
ceased.
Despite these efforts, the census of Galactic \hii\ regions was
clearly incomplete.  The advent of modern high-resolution,
Galactic-scale infrared and radio surveys, coupled with the
unprecedented spectral sensitivity of the NRAO Green Bank Telescope
(GBT), has made the GBT \hii Region Discovery Survey (HRDS)
possible.

\section{TARGET SAMPLE\label{sec:targets}}

We assembled our target list from the following multi-frequency, large
solid angle Galactic surveys: the infrared {\em Spitzer} Galactic
Legacy Infrared Mid-Plane Survey Extraordinaire
\citep[GLIMPSE:][]{benjamin03} and MIPSGAL \citep{carey09}, the NRAO
VLA Galactic Plane Survey made in 21\cm\,\hi and continuum
\citep[VGPS:][]{stil06}, the VLA MAGPIS at 20\cm continuum
\citep{helfand06}, and the NRAO VLA Sky Survey
\citep[NVSS:][]{condon98}.  Our targets were selected by
finding objects that showed spatially coincident 24\,\microns\ {\em
  Spitzer} MIPSGAL and $\sim\,$20\cm continuum emission, either from
the VGPS or the NVSS.  Our method was not automated, but instead
relied on visual inspection of radio and IR emission maps.  We removed
known \hii regions, planetary nebulae (PNe), luminous blue variables,
and supernova remnants (SNRs) from the target sample using the SIMBAD
database.

This criterion is a strong indication that a target is emitting
thermally \citep{haslam87, broadbent89}, and therefore is likely an
\hii\ region or a PNe.  Warm dust absorbs the far ultraviolet
radiation from the exciting star(s) and re-emits in the IR, whereas
plasma ionized by the same star(s) gives rise to free-free thermal
emission at \cm-wavelengths.  It is certainly true that the radio
continuum can result from a mixture of free-free (thermal) and
synchrotron (non-thermal) emission.
\citet{furst87} showed that one can discriminate thermally
from non-thermally emitting objects by using the ratio of the infrared to
radio fluxes.   
The IR/radio flux ratio for \hii regions is typically $\sim$\,100 times 
larger than that for non-thermally emitting SNRs, so it is easy to 
to differentiate between the two by eye.  Our visual inspection of the
IR and radio images should have eliminated all the SNRs from our target
sample. 
Sources showing coincident mid-IR and \cm-wave continuum emission
almost invariably are thermally emitting: 95\% of our sample targets
show hydrogen RRL emission with line to continuum ratios of
$\sim$\,\expo{-1} which together suggests that our targets are
emitting thermally.

\section{GREEN BANK TELESCOPE OBSERVATIONS\label{sec:HRDS}}

\begin{figure}
\vskip 0cm
\centerline{\includegraphics[scale=0.70]{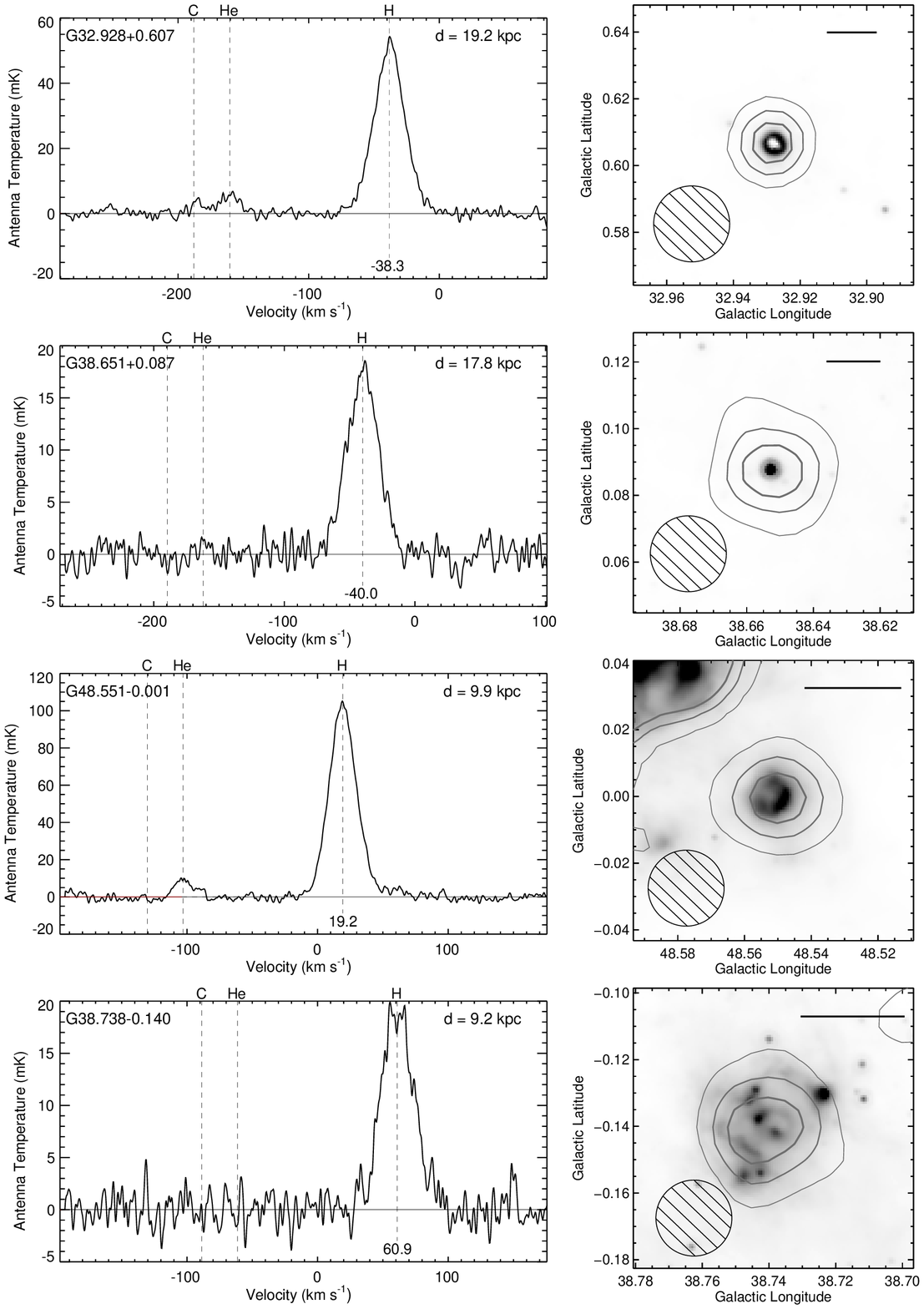}}
\vskip 0cm
\caption{{\bf Left:} GBT HRDS RRL spectra for four new \hii\ regions.  
Shown are the average spectra for seven RRL transitions, \hal{87} to
\hal{93}, smoothed to 1.86\kms\ resolution.  Kinematic distances
to these nebulae, d, were resolved using the RRL velocity and \hi\ 
maps to study \hi\ absorption of the \hii\ region's continuum emission. 
The brightest \hii\ regions show He and C recombination lines. {\bf
  Right:} {\it Spitzer} 24\microns MIPSGAL infrared images for the
same sources, together with contours of VGPS 20\cm continuum emission
(1\arcmin\ resolution).  Contours are drawn at 80\%, 60\%, and 40\% of
the peak emission. All images are 5\arcmin\ squares; each 
scale bar is 5\pc long.  The GBT 
82\arcsec\ (HPBW) beam is shown as a hatched circle.
}
\label{fig:detections}
\end{figure}

Figure~\ref{fig:detections} shows some representative GBT HRDS RRL
detection spectra together with {\em Spitzer} MIPSGAL 24\,\microns\ 
images of the nebulae with contours of the VGPS 20\cm continuum 
emission superimposed.  (The MIPS detector is saturated in the central
region of G32.928+0.607.)  
The sensitivity of the GBT and the power of its Autocorrelation
Spectrometer together made the HRDS possible.  To achieve high
sensitivity, we used techniques pioneered by \citet{balser06}, who
realized that there are 8 H$n\,\alpha$ RRL transitions, \hal{86} to
\hal{93}, that can be measured simultaneously by the GBT with the ACS
at 3\cm wavelength (X-band).  (The \hal{86} transition, however, is 
spectrally compromised by confusing, higher order RRL transitions.)
\citet{balser06} showed that all these transitions can be averaged
(after they are re-sampled onto the same velocity scale) to
significantly improve the RRL signal-to-noise ratio, thus giving an
extremely sensitive X-band H$n\,\alpha$ average nebular spectrum.

This observing technique, coupled with the sensitivity afforded by the
GBT's aperture, gave us unprecedented spectral sensitivity per unit
observing time advantage compared with all previous \cm-wavelength RRL
surveys of Galactic \hii regions.  The vast majority of our detections
took only a single OffOn total power observation.  After Gaussian
smoothing over 5 channels to a resolution of 1.86\kmsch (to be
compared with the $\sim$\,25\kms RRL typical line width), the
r.m.s. noise for a single OffOn observation was typically
$\sim$\,1\mjy.  All details of the HRDS observing and analysis
techniques, together with a catalog of the source properties, are
discussed by Anderson et al. (2010, in preparation).

We also used the Digital Continuum Receiver to measure the free-free
continuum flux for all our HRDS targets.  These show a power law
distribution for fluxes $\gsim120$\,mjy.  Below this the fitted power
law overestimates the observed distribution which implies that the
HRDS is complete to a flux limit of $\gsim120\mjy$.  Using the stellar
fluxes given by \citet{sternberg03}, we estimate that optically thin
\hii\ regions ionized by single O9 stars within the Solar orbit have
flux densities \gsim 120\mjy at 9\ghz.  The HRDS thus should be
finding all such \hii\ regions and, furthermore, is capable of
detecting O3 stars at heliocentric distances of \dsun$\sim$\,20\kpc
\citep{rubin68, anderson09}.

\section{DISCUSSION\label{sec:discuss}}

\begin{figure}
\vskip -5cm
\includegraphics[scale=0.80,angle=0.]{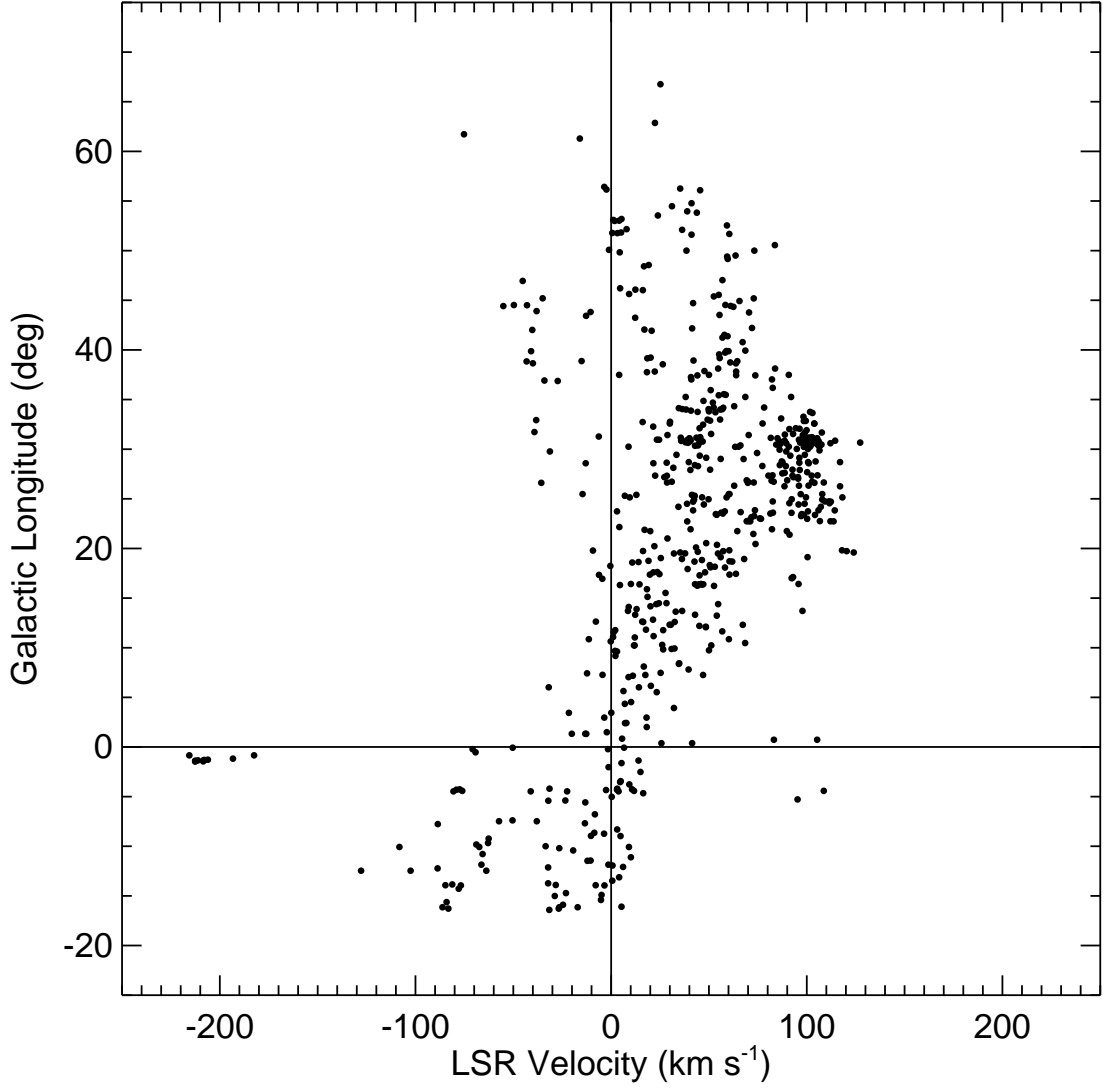}

\caption{Longitude-Velocity distribution of the 602 radio recombination lines
found by the GBT \hii Region Discovery Survey projected onto
the Galactic plane.  
}
\label{fig:LV_250}
\end{figure}

The GBT HRDS has doubled the number of known \hii regions in the
Galactic zone $-16\,\degr \leq \gl \leq 67\,\degr$ and $\vert \gb
\vert \leq 1\,\degr$.
Each discovery gives a position and LSR velocity for the nebula.  We
detected 602 discrete Hydrogen RRL components from 448 target lines of sight,
including $\sim$\,65 infrared bubbles found in the {\em Spitzer} GLIMPSE 
survey.  We found 129 multiple velocity component sources (29\% of the target
sample).
Figure~\ref{fig:LV_250} shows the longitude-velocity distribution of
the HRDS nebulae, projected onto the Galactic plane.

Although the HRDS found 448 \hii regions, the number of physically
distinct nebulae represented by the 602 discrete Hydrogen RRL
components is not well known.  Thermal radio sources can often be
resolved into several apparently physically distinct emission regions,
each having a somewhat different position and RRL velocity.
Furthermore, the most massive star-forming complexes, W\,43 for
example, are extended and fragmented into many sub-clumps of localized
star formation, which together can ionize a very large zone.  Many of
our multiple velocity component targets may be detecting such low
density ionized gas in addition to RRL emission from another,
physically distinct nebula.  Here we follow the convention established
by \citet{lockman89} and assume that each of our 602 RRLs is produced
by a distinct object.  One should keep in mind, however, that the
concept of a ``discrete \hii region'' may not apply to many of the
complex nebulae seen in the inner Galaxy.

Here we assume that all the HRDS sources are \hii regions.  
Anderson et al. (2010, in preparation) show that the level 
of contamination in the HRDS sample by heretofore unknown SNRs,
PNe, luminous blue variables, etc., is very small.  This conclusion 
is based on considerations of galactic structure, scale height, 
RRL line widths, and the RRL line-to-continuum ratios (i.e. nebulae
electron temperatures).  

Here we use the \citet{paladini03} catalog as a proxy for the sample
of previously known \hii regions.  The \citet{paladini03} catalog of
Galactic \hii\ regions is a compilation of 24 Galactic \hii\ region
single-dish, medium resolution ($\sim$\,few arcmin) surveys covering
the entire Galactic plane.  There is, however, no definitive, complete
compilation of all previously detected Galactic \hii regions.  Our
HRDS nebulae are new, previously unknown \hii regions because they are
not listed in the \citet{paladini03} compilation, the
\citet{lockman96} survey, the Ultra-Compact nebulae studied by
\citet{araya02}, \citet{watson03}, and \citet{sewilo04}, nor are they
in the SIMBAD database.

\begin{figure}
\vskip -10cm
\includegraphics[scale=0.85,angle=0.]{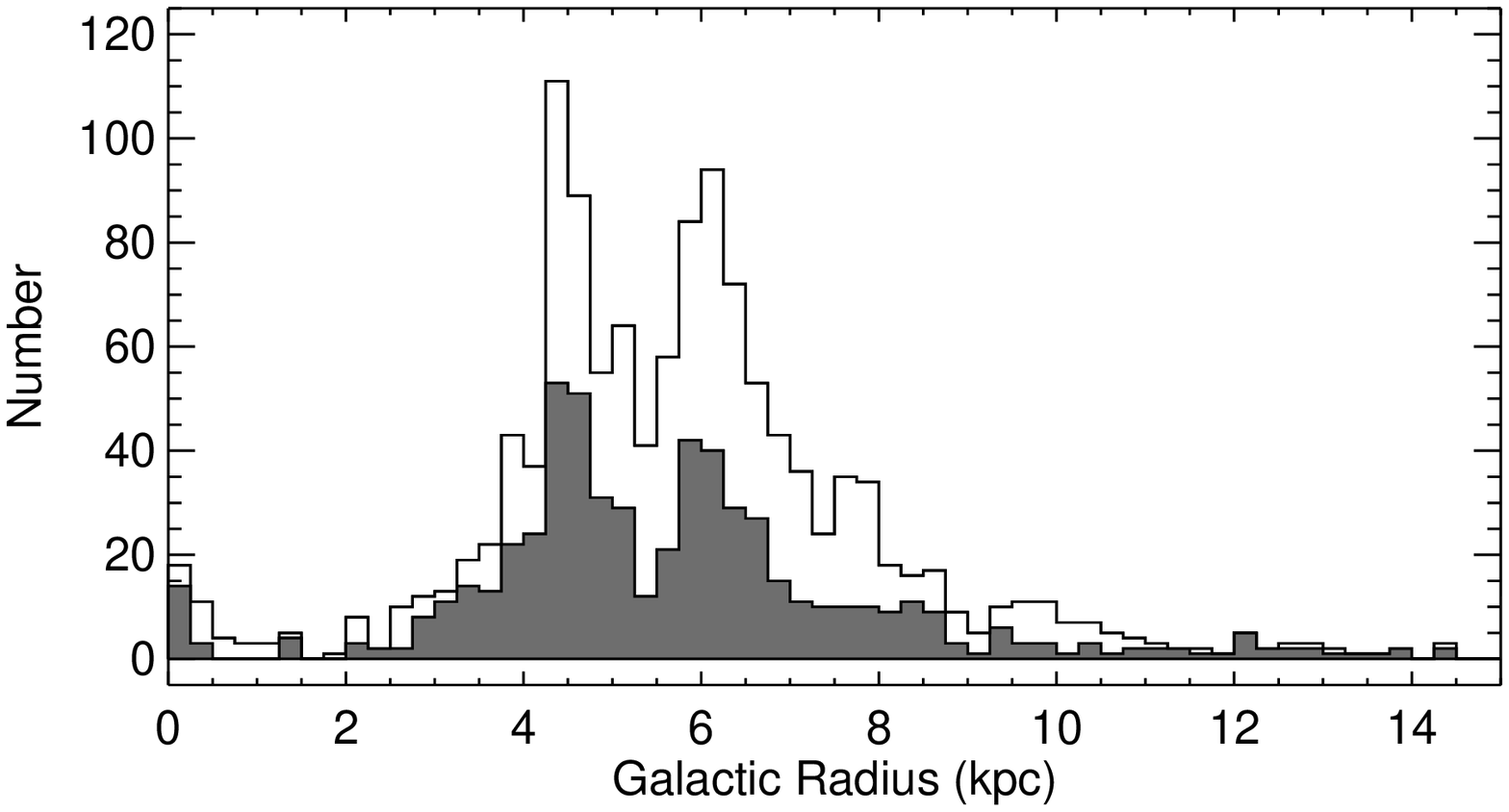}

\caption{Radial distribution of Galactic \hii\ regions.  The Galactocentric
radius of each nebula, \rgal, is derived from the observed RRL LSR
velocity using the \citet{brand86} rotation curve.  The filled histogram
shows the 602 new nebulae found by the GBT HRDS.  The open histogram
shows the distribution of the 1,276 \hii\ regions in the combined 
HRDS and \citet{paladini03} samples.  These histograms are averaged
in Galactic azimuth, yet they show two significant, narrow (\rgal\lsim1\kpc)
peaks at \rgal = 4.25\kpc and 6.0\kpc.
}
\label{fig:RgalHist}
\end{figure}

The LSR velocity of each HRDS \hii region maps into a unique
Galactocentric radius, \rgal.  Using the \citet{brand86} rotation
curve, Figure~\ref{fig:RgalHist} shows the azimuthally averaged
distribution of Galactic \hii regions as a function of \rgal.  The
filled histogram shows the HRDS nebulae; the open histogram is the
total \hii region sample \citep[1276 nebulae: HRDS \&][]{paladini03}.
The distribution of \hii regions across the Galactic disk shows
strong, narrow (\rgal\lsim1\kpc wide) peaks at Galactic radii of 4.25
and 6.00 kpc.  There is still an overall dearth of \hii regions within
4\kpc radius \citep{burton76, lockman81, bronfman00}. In the \rgal = 2
to 4\kpc zone, however, the HRDS has found 75 new nebulae, whereas
\citet{paladini03} list only 56.  The majority of these HRDS RRL
components appear to be associated with the Galactic Bar at \rgal
$\sim$ 3\kpc.  The Galactic Bar and 3 Kpc Arm, however, produce large
streaming motions that occur throughout the inner Galaxy, making all
\rgal determinations that assume circular rotation uncertain inside 4
kpc radius \citep{burton93}.

Together, Figs.~\ref{fig:LV_250} and \ref{fig:RgalHist} show that
the level of PNe contamination in the HRDS sample must be minimal.
Because PNe are an old stellar population their Galactic orbits are
well-mixed.  PNe show, therefore, no structure in their Galactocentric
radial distribution and their Galactic longitude-velocity distribution
is a scatter plot constrained only by velocities permitted by Galactic
rotation.  Any PNe contamination of the HRDS sample must therefore be
very small, otherwise these interlopers would suppress the unambiguous
signal of Galactic structure seen in the HRDS Galactocentric radial
and Longitude-Velocity distributions.
 
Figure~\ref{fig:RingArms} shows that the longitude-velocity
distribution of the new HRDS and previously known \hii region sample
together now give unambiguous evidence for an ordered pattern of
Galactic structure.  Doubling the census makes the contrast of the
\gl--$V$ features striking.  The empty zones in Fig.~\ref{fig:RingArms} are
just as important in this regard as are the features seen. The sample
of 1276 nebulae clearly shows the kinematic signatures of the radial
peaks in the spatial distribution (Fig.~\ref{fig:RgalHist}), a
concentration of nebulae at the end of the Galactic Bar, at
\gl$\sim$\,30\,\degr\, and \vlsr\,$\sim$\,+100\kms \citep{benjamin05,
  churchwell09}, and nebulae located on the kinematic locus of the
3\kpc Arm.

The \hii region Galactocentric radial distribution peaks at 4.3\kpc
and 6.0\kpc have traditionally been associated with the
Scutum-Centaurus and Sagittarius spiral arms.  In the first Galactic
quadrant they imply tangent point longitudes of 30\degr\, and
45\degr, respectively, for \ro\,=\,8.5\kpc.  The \citet{brand86}
rotation curve terminal velocities for these directions are 107\kms
and 65\kms, respectively.  The straight lines in
Fig.~\ref{fig:RingArms} are the solid body loci defined by these
tangent points and terminal velocities.  These loci trace the
over-densities in the \hii region \gl--$V$ distribution for $\gl \geq
0\,\degr$ quite well.  These Northern tangent point longitudes show
evidence for non-circular streaming motions of $\sim$\,10\kms.

Over 50 years after its discovery \citep{vanwoerden57}, the
precise astrophysical nature of the Milky Way's 3 Kpc Arm remains
enigmatic.  The ellipse drawn in Fig.~\ref{fig:RingArms} shows the
locus of the \citet{cohen76} 3 Kpc Arm kinematic expanding ring model
(\rgal=\,4\kpc, $V_{exp}=53\kms$, $V_{rot}=210\kms$).  Discovered by the
very first 21\cm\,\hi surveys, the 3 Kpc Arm also contains considerable
amounts of molecular gas \citep{bania77, bania80, bania86}.  Although
the near side of the Arm, the segment between the Sun and the Galactic
Center, is quite prominent, the far side of the Arm was only recently
discovered by \citet{dame08}.  Because of this spatial symmetry and
its extreme non-circular velocities, the 3 Kpc Arm provides strong
evidence that the Milky Way is a m=2 barred spiral galaxy \citep{fux99}.

Knowing that on-going star formation is occurring in the 3 Kpc Arm may
help constrain theories of its dynamical origin.  The 3 Kpc
Arm lacks large numbers of \hii regions \citep{lockman80,
  lockman81}, but there are some in it \citep{bania80}.
Although the HRDS found a few more nebulae ($\sim$\,10 more in the
near side Arm, see Fig.~\ref{fig:LV_250}), it has not, however, found
a substantial new population of \hii regions along the 3 Kpc Arm
\gl--$V$ locus.  The far side of the Arm, in the $(\gl\leq0\,\degr,
\vlsr\geq0\kms)$ quadrant, remains almost devoid of nebulae along the
Arm locus.  
Because the velocities predicted by the expanding ring model will
blend with normal Galactic rotation velocities in the $(\gl\geq0\,\degr, 
\vlsr\geq0\kms)$ and $(\gl\leq0\,\degr, \vlsr\leq0\kms)$ quadrants, the \hii 
region clusterings seen along the \citet{cohen76} ring locus are 
difficult to interpret.

We were able to determine the distances to G$38.737-0.140$ and
G$48.551-0.001$ in Fig.~\ref{fig:detections} by using \hi
emission/absorption experiments to resolve the kinematic distance
ambiguity \citep[As][did for the sample of previously known first 
Galactic quadrant \hii regions.]{anderson09a} Each nebula is at the
far kinematic distance.  The majority of our HRDS \hii regions are
unresolved with our 82\arcsec\ survey resolution.  We are finding that
many of these small angular diameter nebulae lie at the far kinematic
distance.  Earlier RRL surveys missed these nebulae because their weak
continuum made them poor target choices.

The HRDS found 25 first quadrant nebulae with negative LSR velocities.  
In the first Galactic longitude quadrant, a negative RRL LSR velocity
unambiguously places the \hii region beyond the Solar orbit, at large
distances from the Sun, \dsun \gsim 12\kpc, in the outer Galactic
disk, \rgal \gsim 9\kpc.  Prior to the GBT HRDS, there were only two
negative-velocity \hii\ regions known in the $18\,\degr \leq \gl \leq
55\,\degr$, $\vert \gb \vert \leq 1\,\degr$ zone.  The newly discovered
Fig.~1 sources ${\rm G38.651+0.087}$ at $-$40.0\kms and ${\rm G32.928+0.607}$ at
$-$38.3\kms already match the size of the previous census.  
The \gl--$V$ distribution of these negative velocity sources shows good 
agreement with \co\ maps made by \citet{dame01}.  This region in 
\gl,$V$-space has traditionally been termed the ``Outer Arm''. 

Because of their location in the critical region beyond the Solar
orbit at $\rgal\,\sim\,9-12$\kpc, these nebulae will provide new GCE
constraints.  The HRDS and follow-up GBT observations will allow us to
derive the nebular electron temperature, \te, and helium abundances (Y
= \he4/H).  Because metals are the main coolants in the photo-ionized
gas, both \te and Y are directly related to the distribution of heavy
elements in the Milky Way.  There are relatively few \hii\ regions
with accurately derived \te values, especially at the critical
\rgal\,$\sim$\,10 \kpc region.  In the first Galactic quadrant, our 25
new HDRS nebulae can increase the \te sample size by a factor of 10.

We detected RRL emission from 65 \hii regions that are surrounded by
mid-infrared bubbles.  The {\em Spitzer} GLIMPSE survey found almost
600 objects in the inner Galaxy that have a ring-shaped morphology
\citep{churchwell06, churchwell07}.  These are presumably
bubbles that are viewed in projection.  These objects have simple
morphologies and ``swept-up'' neutral material; they may be sites of
triggered star formation \citep[see, e.g.,][]{deharveng09}.  
\cite{churchwell06} argue that $\sim$\,75\% of GLIMPSE bubbles are
caused by B-stars without detectable \hii regions.  Because of the  
large number of bubbles that we find enclosing HRDS sources,
we speculate that nearly all GLIMPSE bubbles are caused by 
\hii regions. 

\section{SUMMARY\label{sec:summary}}

\begin{figure}
\vskip -5cm
\includegraphics[scale=0.8,angle=0.]{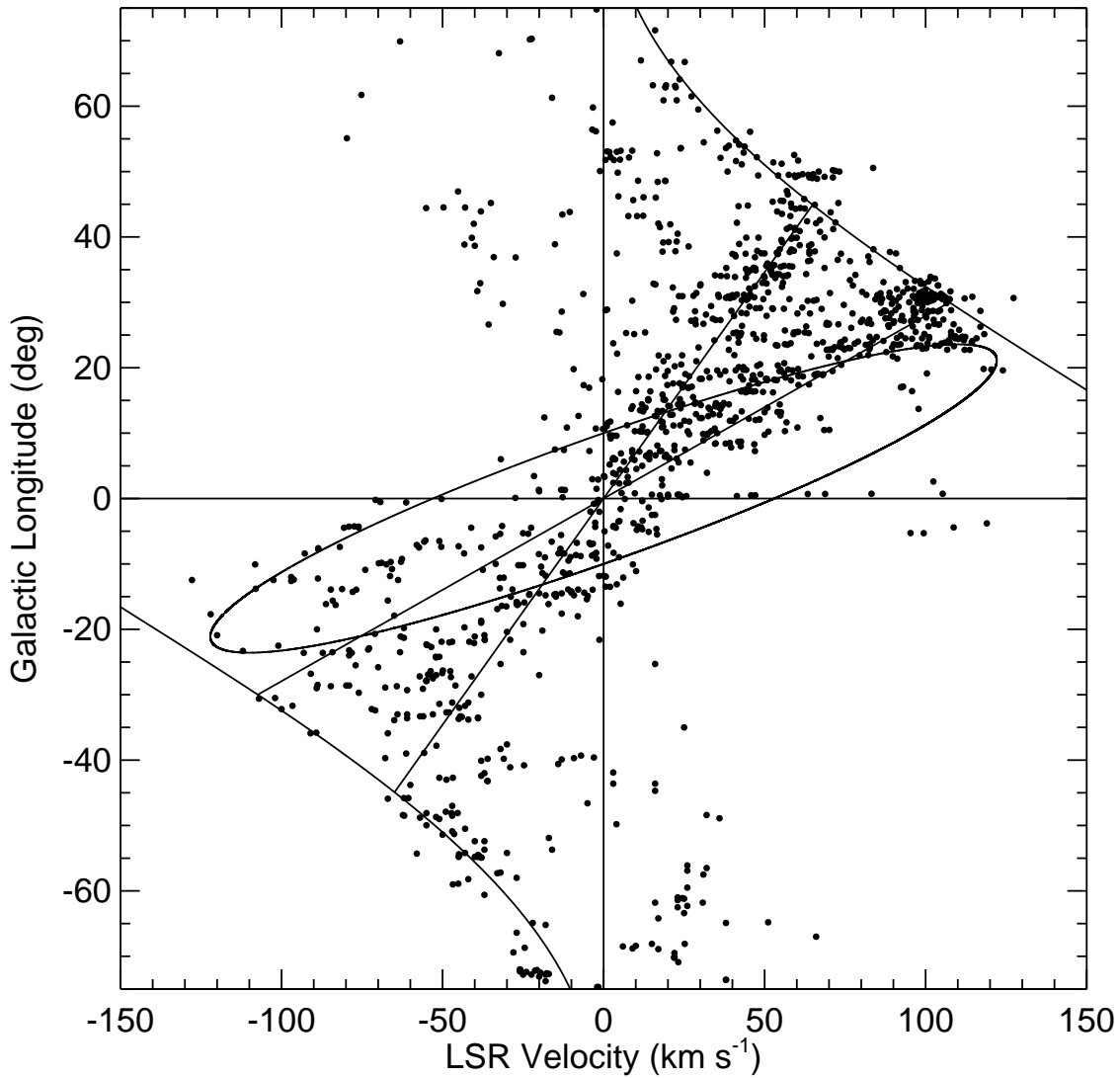}

\caption{Longitude-Velocity distribution of Galactic \hii\ 
regions (the combined GBT HRDS \& Paladini et al. 2003 samples) 
with kinematic Galactic structure loci.  For clarity we 
have not plotted nebulae in the Nuclear Disk at large LSR velocities, 
$\vert V \vert \geq 150\kms$.  Shown are the 
\lv loci for: (1) the circular rotation solid bodies for the 
peaks in the \hii region Galactocentric radial distribution 
(straight lines: Fig.~\ref{fig:RgalHist}); (2) the \citet{cohen76} 
3~Kpc Arm kinematic ring model; and (3) the circular rotation 
terminal velocity derived from the \citet{brand86} rotation curve.
}
\label{fig:RingArms}
\end{figure}

The advent of modern high-resolution, Galactic-scale infrared and
radio surveys, e.g.  {\em Spitzer} GLIMPSE/MIPSGAL and the VLA VGPS,
coupled with the unprecedented spectral sensitivity of the NRAO Green
Bank Telescope allowed us to make a major new discovery survey of
Galactic \hii\ regions. The GBT HRDS has doubled the number of known
\hii regions in the Galactic zone $-16\,\degr \leq \gl \leq 67\,\degr$
and $\vert \gb \vert \leq 1\,\degr$.  The census of \hii regions, when
enhanced by the HRDS, now shows in this Galactic zone a
longitude-velocity distribution that gives unambiguous evidence for
Galactic structure, including the kinematic signatures of peaks in the
radial spatial distribution of nebulae, a concentration of nebulae at
the end of the Galactic Bar, and nebulae located on the kinematic
locus of the 3\kpc Arm.  Doubling the Galactic \hii region census
makes the contrast of the \gl--$V$ features striking.  The empty zones
in Fig.~\ref{fig:RingArms} are just as important in this regard as
are the concentrations of nebulae in \gl--$V$ space.

We found 25 new nebulae located beyond the Solar orbit, at large
distances from the Sun, \dsun \gsim 12\kpc, in the outer Galactic
disk, \rgal \gsim 10\kpc.  Because of their location, these nebulae
will be important for future studies of the radial metallicity
gradient in the Galaxy.  Many of our new nebulae are seen as bubbles
in {\em Spitzer} GLIMPSE images.  We found 65 such objects and
speculate that nearly all the {\em Spitzer} GLIMPSE mid-IR identified
bubbles are \hii regions.

The HRDS nebular distances will be determined by using \hi
emission/absorption experiments to resolve the kinematic distance
ambiguity \citep[see][]{anderson09a}. Because we can detect all
nebulae inside the Solar orbit that are ionized by O-stars, the GBT
HRDS sources, when combined with existing \hii region catalogs, will
provide a more complete census of Galactic \hii regions with known
distances and physical properties, which is the fundamental database
needed for ISM evolution studies of the Molecular
Cloud/\hii\ Region/Star Cluster/Supernova Bubble life-cycle\/.  This
will enhance our ability to study Galactic structure using spatial
distributions and to constrain Galactic chemical evolution using
spatial patterns of nebular metallicity.

\acknowledgments We thank those visionaries who came before us for the
support to continue this sort of fundamental survey science.  The
National Radio Astronomy Observatory is a facility of the National
Science Foundation operated under cooperative agreement by Associated
Universities, Inc.  LDA was partially supported by the NSF through
GSSP awards 08-0030 and 09-005 from the NRAO.

\clearpage

\clearpage

\clearpage

\clearpage

\end{document}